\documentstyle[aps,epsf]{revtex}
\draft
\begin{document}

\title{Amine Functionalized Cross-Linked Hybrid\\
Full Color Phosphors Lacking Metal Activator Ions}

\author{L. D. Carlos$^{(1)}$, R. A. S\'a Ferreira$^{(1)}$,
V. De Zea Bermudez$^{(2)}$, \\ and Sidney J.L. Ribeiro$^{(3)}$}

\address{$^{(1)}$Departamento de F\'{\i}sica, Universidade de Aveiro,
3810 Aveiro, Portugal}

\address{$^{(2)}$ Sec\c c\~ao de Qu\'{\i}mica, Universidade de Tr\'as-os Montes e Alto Douro,
Quinta de Prados, \\
Apartado 202, 5001 Vila Real Codex, Portugal}

\address{$^{(3)}$Instituto de Qu\'{\i}mica - UNESP, C.P. 355, 14800-900,
Araraquara-SP Brazil}

\date{\today}
\maketitle

\begin{abstract}
White light tunable phosphors are obtained from a sol-gel derived
siliceous network to which oxyethylene units are covalently
grafted by means of urea or urethane cross-linkages. The bright
photoluminescence is ascribed to an unusual convolution of
distinct emissions originated in the NH groups of the urea or
urethane bridges and in the siliceous nanodomains. The lifetimes
of these emissions show a remarkable temperature dependence of
approximately six orders of magnitude between $\approx 220$ and
$300$ K associated with the hybrid´s glass transition temperature.
This set of xerogels is one of the most surprising classes of full
color emitters lacking metal activator ions.
\end{abstract}
\newpage
\section{Introduction}

The synthesis of a wide range of novel materials by the sol-gel
via has attracted considerable scientific interest in recent
years\cite{Brinker,Pope,Reisfeld,Judeinstein,Sanchez}. In
particular, the advantages (large variety and high purity of
suitable precursors) of the rich chemistry of silicon-based
networks have been employed to synthesize stable and efficient
white light photoluminescent organic/inorganic hybrids lacking
metal activator ions
\cite{Sailor,Carlos98,Bermudez98,Lianos98,Lianos98a,Carlos99,Bermudez99,Carlos99a}.
The outgrowth of full color displays, cheaper and less aggressive
to the global environment, which could replace with substantial
advantages the metal emissive centers currently used is one of the
main challenging tasks for the next generation of flat panel
display systems and lighting technologies. This aspect is well
illustrated by the impressive impact of light emitting diodes
based on conjugated polymers recently developed\cite{Friend,Cao}.
In spite of the potential technological relevance of the white
light emission obtained from organically modified silicates, most
of the research has been focused on the photoluminescence (PL)
features of lanthanide-based hybrid
matrices\cite{Reisfeld,Judeinstein,Sanchez,Carlos98,Costa,Ribeiro,Lianos99,Carlos99c,Carlos00},
seeking essentially to take advantage of the high chromaticity and
long-excited lifetime characteristic of the metal ions, and of the
improvement on the emission features and mechanical stability of
the samples (compared to those of conventional silica gel hosts).
As the hybrid matrices have been basically considered as an
optically inert framework, only a limited amount of work has been
directed to the study of their light emission
features\cite{Sailor,Carlos98,Bermudez98,Lianos98,Lianos98a,Carlos99,Bermudez99,Carlos99a}.
Since the publication of Canham's paper in 1990, demonstrating
efficient tunable room temperature PL from porous silicon
layers\cite{Canham}, the interest in light-emitting silicon-based
materials has been significantly aroused \cite{Wilson}. The
development of an optically efficient silicon-compatible material
permitting optical and electronic devices to be completely
integrated on a silicon wafer will certainly provoke a pronounced
impact on displays, communications, computers and related
technologies\cite{Hirschman}.

In this report, we discuss the origin of the PL features of stable
and environmentally "friendly" innovative sol-gel derived hybrids
containing $\text{OCH}_2\text{CH}_2$ repeat units grafted onto a
siliceous backbone through urea, -NHC(=O)NH-, or urethane,
-NHC(=O)O-, bridges. Since the synthesis of the former series of
xerogels was first reported \cite{Armand,Bermudez92}, several
papers describing their PL features have recently appeared
\cite{Carlos98,Bermudez98,Carlos99,Bermudez99,Carlos99a,Ribeiro,Lianos99,Carlos99c,Carlos00}.
Yet, both the identification of the chemical species responsible
for the white light emission and the characterization of the PL
mechanism still remain controversial issues. We demonstrate here
that the white light PL results from an unusual convolution of a
longer-lived emission ($10^{-1}$ s below $220$ K) originated in
the NH groups of the urea or urethane bridges with shorter-lived
electron-hole recombinations ($10^{-3}$ s below $220$ K) that
occur in the nanometer-sized siliceous clusters. Moreover, a
remarkable temperature dependence of the two lifetimes is observed
around the glass transition temperature, $T_g$, of the hybrids
($T_g\approx 220$ K). The longer-lived lifetime, for instance,
steeply decreases to  $\approx 10^{-6}-10^{-7}$ s, when the
temperature is increased from $200-220$ K to room temperature.
\section{Experimental}

A covalent cross-link between the
3-isocyanatepropyltriethoxysilane precursor, ICPTES (Fluka) and
the oligopolyoxyethylene chains was formed by reacting the
terminal amino groups of doubly functional amines
($\alpha$,$\omega$-diaminepoly(oxyethylene-co-oxypropylene),
Fluka) or the OH groups of PEG (Aldrich), with the isocyanate
group of the precursor in tetrahydrofuran (Merck) at room
temperature and about $350$ K, respectively. A urea- or
urethane-based hybrid precursor was thus obtained
[7,8,11-13,24,25]. Three diamines, commercially designated as
Jeffamine ED-2001®, Jeffamine ED-900®, and Jeffamine ED-600®, and
PEG with different molecular weights are used. In the second stage
of the synthesis process, the resulting hybrids were formed,
adding a mixture of ethanol and water to the precursor solutions
(molar proportion $1$ ICPTES: $4 \,
\text{CH}_3\text{CH}_2\text{OH}$: $1.5 \, \text{H}_2\text{O}$).
The mixture was stirred in a sealed flash for $30$ min and then
cast into a Teflon mold and left in a fume cupboard for $24$ h.
After a few hours, gelation occurs and the mold was transferred to
an oven at $310$ K for a period of $7$ days. The samples were then
aged for $3$ weeks at about $350$ K to form white or yellowish
amorphous and transparent monoliths, thermally stable up to $523$
K that underwent bright PL. The hybrids have been classed,
respectively, as di-ureasils, U(Y), and di-urethanesils, Ut(Y'),
where Y=$600$, $900$, $2000$ and Y'=$300$, $2000$ denote the
average molecular weight of the starting diamine or of the PEG
used.

The emission spectra were recorded ($14-300$ K) under continuous
excitation with a 150 W xenon arc lamp coupled to a $0.25$ m
monochromator (Kratos GM-252) or an Ar ion laser (multi-line
excitation $331.1$ and $351.4$ nm, $45$ mW). A pulsed Xe arc lamp
($5$ mJ/pulse, $3$ µs bandwidth) and a SPEX 1934 C phosphorimeter
are used in the time-resolved measurements. All spectra were
corrected for the spectral response of the monochromator (1704
Spex) and the photomultiplier (Hamamatsu R928).
\section{Results and Discussion}

\begin{figure}[tbp]
\begin {center}
\leavevmode
\hbox{%
\epsfxsize 4.0in \epsfbox{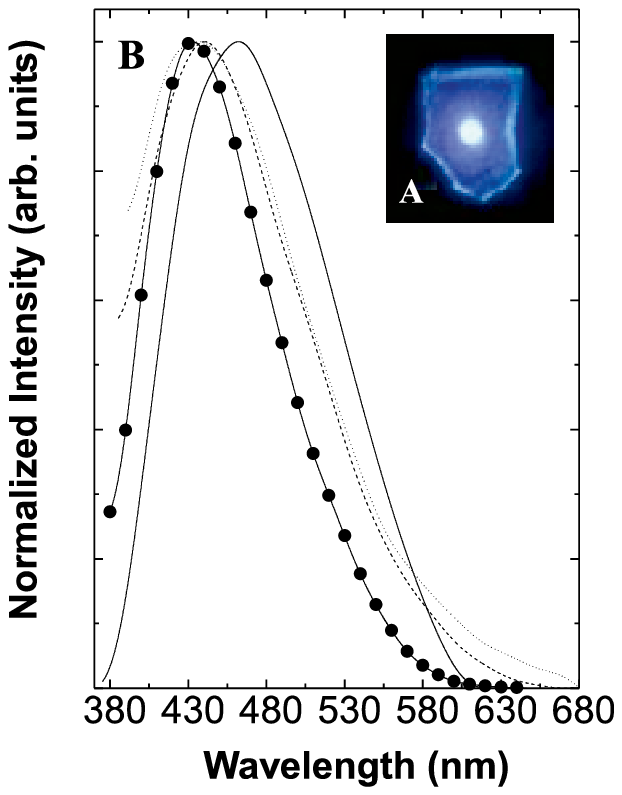}}
\end{center}
\caption {A: A photograph showing the white light emission of
U(2000) at 298 K under UV excitation (Ar ion laser). The diameter
of the emitting area is $\approx 3$ mm$^2$. B: PL spectra (365 nm
excitation at 298 K) of representative amine-based
functionalities, precursors and di-ureasil and di-urethanesil
hybrids: Jeffamine ED-2001® (dashed curve), non-hydrolyzed
Ut(2000) precursor (dotted curve), U(2000) (solid curve), and
Ut(2000) (solid curve with circles).}\label{fig1}
\end{figure}

The di-ureasil and di-urethanesil hybrids display a bright white
light room-temperature PL (Fig.\ \ref{fig1}A), which is similar to
the one detected for the corresponding diamines and the
non-hydrolized organic/inorganic precursors (Fig.\ \ref{fig1}B).
The emission of the diamines and the non-hydrolyzed di-ureasil
precursors are identical demonstrating hereby that the formation
of urea cross-linkages does not alter any of the PL features (Fig.
1B). After polycondensation and the building up of the inorganic
network, a red shift of the PL maximum intensity is clearly
observed ($\approx 1061$ cm$^{-1}$ for U(2000), at an excitation
of $365$ nm, Fig.\ \ref{fig1}B). Yet, comparing the PL spectra for
the two series of hybrids, the spectrum of Ut(2000) is
blue-shifted with respect to the U(2000) one ($\approx 1470$
cm$^{-1}$ at an excitation of $365$ nm, Fig.\ \ref{fig1}B). When
the excitation wavelength increases from $325$ to $420$ nm, the
broad emission of the Jeffamines and urea- and urethane-based
precursors is strongly shifted towards the red (not shown). This
dependence is identical to the already reported behavior of the
di-ureasils and di-urethanesils, permitting thereby the remarkably
easy fine tuning of the hybrids color by varying only the
excitation wavelength\cite{Carlos99,Carlos99a}. These results
clearly indicate that the PL of the hybrids receives a major
contribution from the NH groups. As no emission is detected from
pure PEG, the PL must be related with the lone pair of electrons
of the NH groups present in the diamines and in the non-hydrolized
precursors, a claim which is in total disagreement with one
reported very recently for similar hybrids\cite{Lianos99}.

\begin{figure}[tbp]
\begin {center}
\leavevmode
\hbox{%
\epsfxsize 4.0in \epsfbox{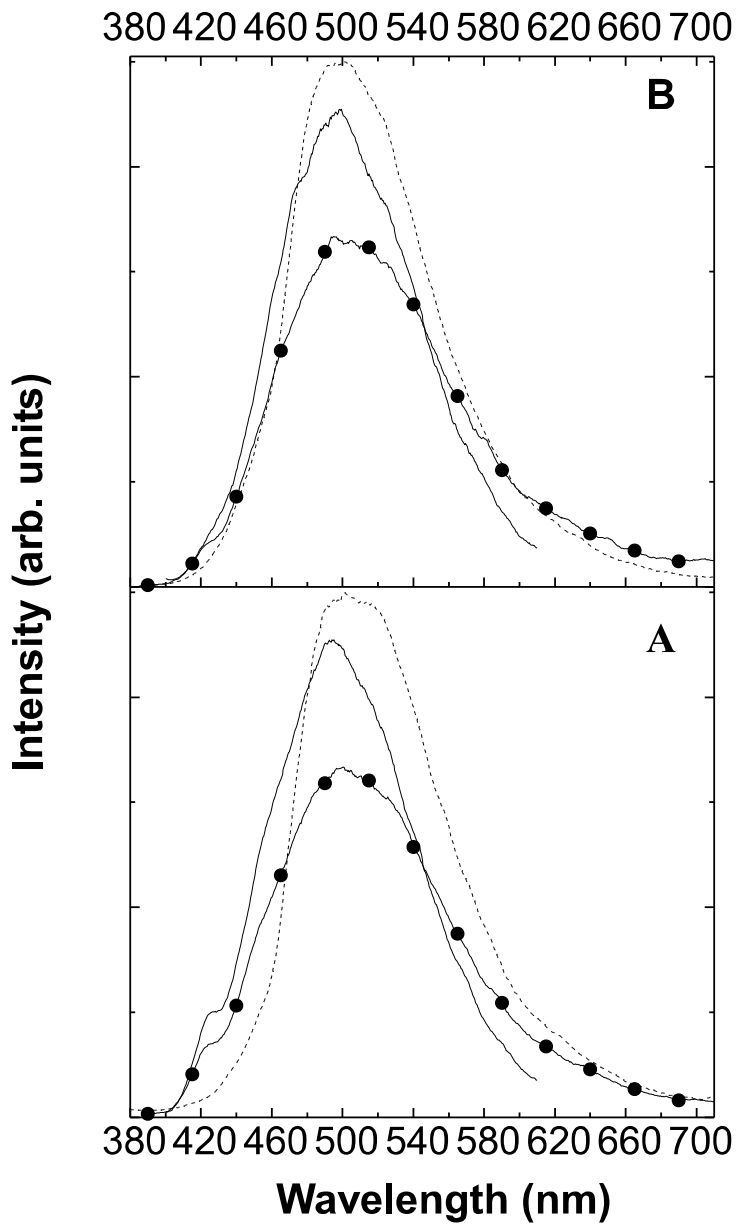}}
\end{center}
\caption {Time-resolved spectra ($365$ nm excitation at $14$ K) of
Jeffamine ED-2001 (dashed curve), U(2000) (solid curve), and
Ut(2000) (solid curve with circles) measured at a fixed
acquisition window, $10$ ms, and different delay-times, A: $0.08$
ms, B: $5$ ms.}\label{fig2}
\end{figure}

\begin{figure}[tbp]
\begin {center}
\leavevmode
\hbox{%
\epsfxsize 4.0in \epsfbox{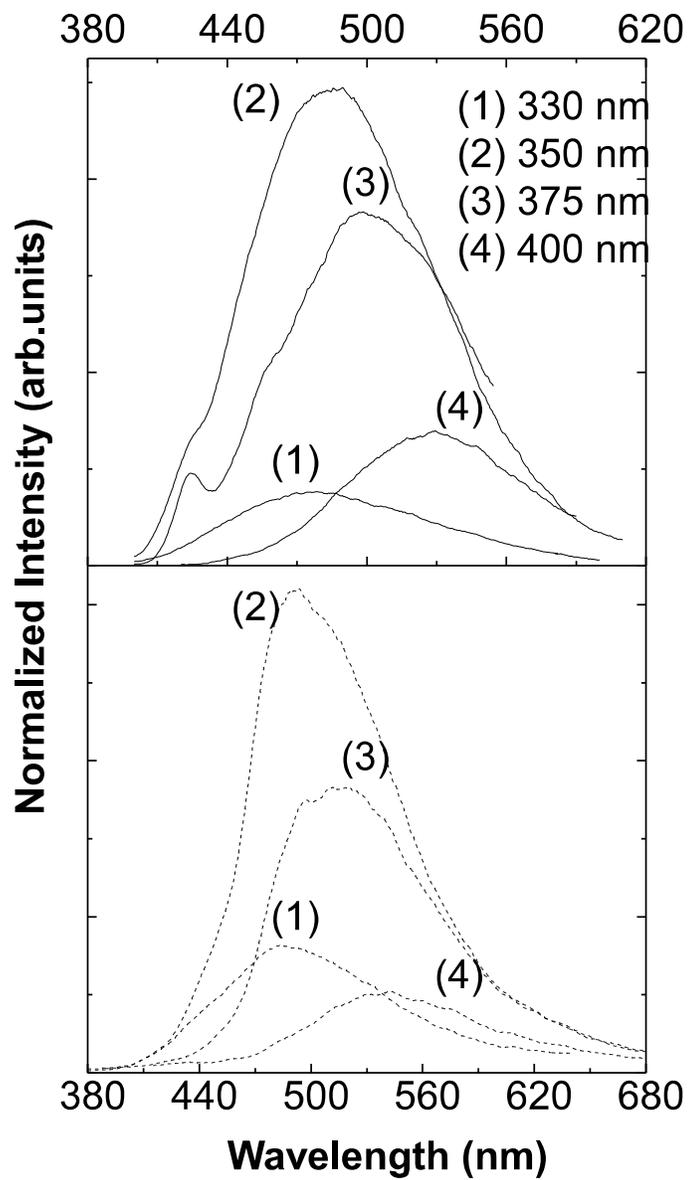}}
\end{center}
\caption {Time-resolved spectra ($0.08$ ms delay-time, $10$ ms
acquisition window at $14$ K) for U(2000) (solid curves) and
Jeffamine ED-2001 (dashed curves) obtained for different
excitation wavelengths.}\label{fig3}
\end{figure}

The differences observed between the steady-state emission of the
hybrids and that of the respective precursors are well established
by time resolved spectroscopy measurements. For delay-times
smaller than $5$ ms, while the U(2000) and Ut(2000) spectra
unambiguously display two bands in the blue and purplish-blue
spectral regions\cite{Carlos99,Carlos99a}, only a single large
broad band is seen in the spectrum of Jeffamine ED-2001, Fig.\
\ref{fig2}. For the diamines and the non-hydrolized precursors
there is no evidence of the presence of the band that clearly
appears centered at about $\approx 425$ nm in the time-resolved
spectra of all the five hybrids. Besides having different
intrinsic time scales, the two bands in the di-ureasil and
di-urethanesil spectra also exhibit a distinct behavior when the
excitation wavelength is increased from $325$ to $420$ nm (Fig.\
\ref{fig3}). The blue band is evident over the whole excitation
wavelength range, the location of its intensity maximum strongly
depending on the excitation wavelength. The purplish-blue band,
which displays no energetic position dependence with the
excitation wavelength, could be detected only between $350$ and
$375$ nm (Fig.\ \ref{fig3}).

\begin{figure}[tbp]
\begin {center}
\leavevmode
\hbox{%
\epsfxsize 4.0in \epsfbox{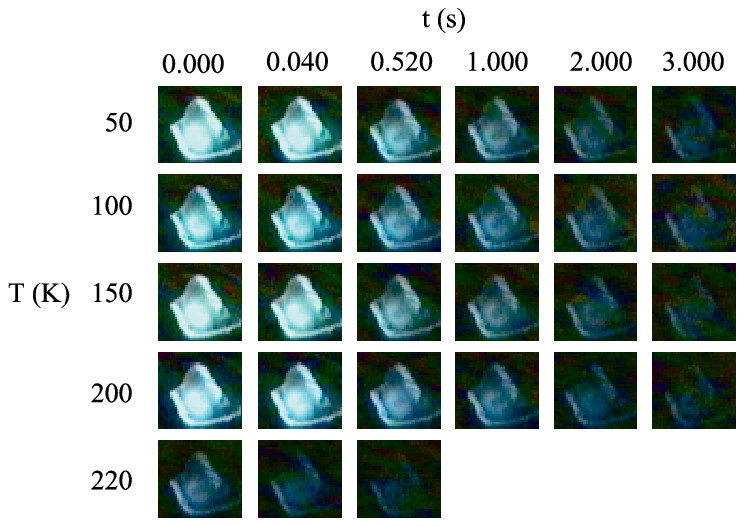}}
\end{center}
\caption {A set of frames showing the real time dependence
($14-220$ K) of the U(2000) white light emission during $3$ s
after the Ar laser beam was turned off. Note that the emission
color is slightly distorted by the video camera. The real color is
more "whitish-blue", Fig. 1A.}\label{fig4}
\end{figure}

For all the five hybrids, the $14$ K blue and purplish-blue band
lifetimes are $\approx 160.0$ and $\approx 3.5$ ms, respectively
(errors within $3-5$\%), in the whole range of excitation
wavelengths used. The observed lifetime for the Jeffamines, at the
same temperature and excitation wavelengths, is $\approx 120.0$ ms
($\pm 3-5$\%). By increasing the temperature up to $220$ K, the
lifetime of the purplish-blue band displays a typical Arrhenius
behavior with temperature (not shown). On the contrary, the
long-lived lifetime of the hybrids and of the diamines oscillates
around the low temperature value, therefore displaying no
significant temperature dependence within this range (not
shown).This is another apposite argument supporting the above
assumption that the blue component of the xerogels emission is
fundamentally related to the NH groups of the cross-links. A
noteworthy aspect in these measurements is the abrupt decrease of
the lifetimes above $200-220$ K, nicely illustrated by the set of
real time frames of Fig.\ \ref{fig4}. For temperatures higher than
$\approx 220$ K, the lifetimes lie in a time scale smaller than
$10^{-5}$ s, being around $10^{-8}$ s at room
temperature\cite{Lianos99}.

The PL characterization of the Jeffamines, urea- and
urethane-based non-hydrolized precursors, and the di-ureasil and
di-urethanesil hybrids furnish persuasive arguments that the full
color emission must involve the NH groups. Furthermore, we have
presented strong evidence that grafting the NH$_2$ or OH groups to
the isocyanate group - with the subsequent formation of urea or
urethane cross-linkages, respectively - does not alter
significantly the emission features. Nevertheless, a new band with
completely distinct features is detected, after gelation. Based on
small-angle X-ray scattering (SAXS) results the morphologies of
the di-ureasil and di-urethanesil hybrids were depicted as a
diphasic structure constituted of dispersed and spatially
correlated siliceous nanoclusters embedded in the polymer matrix
and located at the ends of the organic
chains\cite{Carlos99,Karim}. As a result, the purplish-blue
component must be unquestionably related to the silicon-rich
nanodomains. This can be further confirmed by altering the
conditions at which the hydrolysis and condensation reactions
occur, thus modifying the local structures of those siliceous
nanoregions. For the di-ureasils and di-urethanesils reported
here, $^{29}$Si magic-angle spinning nuclear magnetic ressonance
spectroscopy (MAS NMR) has established the presence of essentially
two structures, (SiO)3Si(CH2)- and (SiO)2Si(OH)(CH2)-, surrounding
the Si atoms\cite{Carlos99a}. A new U(600) di-ureasil sample was
synthesized using a different
ICPTES:$\text{CH}_3\text{CH}_2\text{OH}$:$\text{H}_2\text{O}$
ratio, which alters both the condensation degree and the relative
proportions between the two local Si surroundings, as $^{29}$Si
MAS NMR results indicate (not shown). The $14$ K time-resolved
spectra for that hybrid displays a long-lived blue band similar to
the one observed in all the other samples (not shown). However, a
new short-lived band, centered around $450$ nm, is detected on the
purplish-blue region, supporting the suggested influence of the
siliceous nanoclusters on the full color emission detected.

Recently, the origin of the strong PL detected in an
organic/inorganic hybrid very similar to the U(900) di-ureasil -
the only differences in the two synthesis procedures being the use
of NH$_4$F as a catalyst and a distinct Si:H$_2$O molar proportion
- was explained in terms of the introduction of carbon impurities
in the -Si-O-Si- network during hydrolysis-condensation
reactions\cite{Lianos99}. A similar carbon defect has been
previously assumed to be responsible for the PL of hybrid matrices
made from the reaction of tetramethoxysilane and tetraethoxysilane
with a variety of organic carboxylic acids\cite{Sailor}. The
creation of a carbon substitutional silicon defect needs, however,
a heating process at temperatures above $520$ K, at
least\cite{Sailor}. Moreover, water-soluble hybrids generated from
the reaction of (3-aminopropyl)triethoxysilane with a variety of
organic carboxylic acids do not require any heating process above
$330$ K to be luminescent. Accordingly, the luminescent species
must be different than carbon defects and the white light emission
of APTES-derived hybrids was related to the presence of amide
functionalities\cite{Sailor}. All the PL results of the
di-ureasils and di-urethanesils are in agreement with that
hypothesis, stressing, therefore, the significant role of the
lone-pair of the NH groups in their white light emission.

A final comment regarding the remarkable and drastic decrease of
$5-6$ orders of magnitude in the lifetimes, at temperatures
$\approx 220$ K. This critical temperature region concerning the
lifetime determination is exactly the one associated to the T$_g$
of the $\text{OCH}_2\text{CH}_2$ repeat units. Below T$_g$, the
emitter centers related to the NH groups are quenched in a
particular configuration, decreasing, therefore, the non-radiative
deexcitation paths induced by the segmental mobility of the
polymer chains. When the temperature is raised above $\approx 220$
K there is an obvious enlargement in the segmental mobility of the
polymer chains (and consequently in the NH$_2$ chain end-groups)
with the subsequent increase in the possible non-radiative
deexcitation channels.
\section{Conclusions}
The interest in organic-inorganic hybrids is basically associated
with the extraordinary implications for the tailoring of novel
multi-functional advanced materials induced by the mixture at the
nanosize level of organic and inorganic components in a single
material. The synergy of that combination and the particular role
of the inner interfaces open up exciting new areas in materials
science and related technologies. The bright full color room
temperature PL reported here being an unforeseen example that
nicely illustrates the effects on the optical features of hybrid
materials that could be obtained by the grafting of organic and
inorganic components in a single material. The surprising and
unusual convolution of distinct emissions originated in the NH
groups and in the nanometer-sized siliceous domains, the ability
to tune those emissions to colors across the chromaticity diagram,
and the remarkable temperature dependence ($220-300$ K) of their
lifetimes are innovative features emphasizing the potential of
these hybrids on display silicon-compatible technologies requiring
red, green and blue devices.
\section*{ACKNOWLEDGMENTS}
We thank A. L. L. Videira for helpful discussions and critical
reviews of the manuscript. The financial support from Portuguese,
FCT, (PRAXIS/P/CTM/13175/98, 2/2.1/ FIS/302/94, BD/18404/98) and
Brazilian, FAPESP, agencies is gratefully acknowledged.

\begin{references}

\bibitem[1]{Brinker} C. J. Brinker and G. W.
Scherer {\it Sol-gel Science: The Physics and Chemistry of Sol-Gel
Processing}, (Academic Press, San Diego, CA, 1990).

\bibitem[2]{Pope} E. J. A. Pope, S. Sakka, L. S. Klein {\it Sol-gel Science and
Technology}, Ceramic Transactions (The American Ceramic Society,
Westerville, 1995), Vol. 55.

\bibitem[3]{Reisfeld}R. Reisfeld and C. K. J\o rgensen, in {\it Chemistry,
Spectroscopy and Applications of Sol-Gel Glasses}, R. Reisfeld and
C. K. J\o rgensen, Eds. (Springer-Verlag, Berlin, 1991), p. 207.

\bibitem[4]{Judeinstein}P. Judeinstein and C. Sanchez, {\it J. Mater.
Chem.} {\bf 6}, 511 (1996), and references therein.

\bibitem[5]{Sanchez}C. Sanchez, F. Ribot, B. Lebeau {\it ibid.}
{\bf 9}, 35 (1999), and references therein.

\bibitem[6]{Sailor}W. H. Green, K. P. Le, J. Grey, T. T. Au, M.
J. Sailor, {\it Science} {\bf 276}, 1826 (1997).

\bibitem[7]{Carlos98}L. D. Carlos, V. De Zea Bermudez,
M. C. Duarte, M. M. Silva, C. J. Silva, M. J. Smith, M. Assun\c
c\~ao, and L. Alc\'acer, in {\it Physics and Chemistry of
Luminescent Materials VI}, C. Ronda and T. Welker, Eds.
(Electrochemical Soc. Proc., San Francisco, 1998), Vol. 97-29, p.
352.

\bibitem[8]{Bermudez98} V. De Zea Bermudez, L. D. Carlos, M. C. Duarte,
M. M. Silva, C. J. Silva, M. J. Smith, M. Assun\c c\~ao, and L.
Alc\'acer, {\it J. Alloys and Compounds} {\bf 275-277}, 21 (1998).

\bibitem[9]{Lianos98}V. Bekiari and P. Lianos, {\it Chem. Mater.} {\bf
10}, 3777 (1998).

\bibitem[10]{Lianos98a}V. Bekiari and P. Lianos, {\it Langmuir} {\bf 14}, 3459 (1998).

\bibitem[11]{Carlos99}L. D. Carlos, V. De Zea Bermudez, R. A.
S\'a Ferreira, L. Marques, M. Assun\c c\~ao, {\it Chem. Mater.}
{\bf 11}, 581 (1999).

\bibitem[12]{Bermudez99} V. De Zea Bermudez, L. D. Carlos, L.
Alc\'acer, {\it Chem. Mater.} {\bf 11}, 581 (1999).

\bibitem[13]{Carlos99a}L. D. Carlos, R. A. S\'a Ferreira, I. Orion,
V. de Zea Bermudez, J. Rocha, {\it J. Luminescence}, in press.

\bibitem[14]{Friend}R. H. Friend, R. W. Gymer, A. B. Holmes, J. H. Burroughes, R. N. Marks,
C. Taliani, D. D. C. Bradley, D. A. Dos Santos, J. L. Bredas, M.
Logdlund, W. R. Salaneck, {\it Nature} {\bf 397} 121 (1999) and
references therein.

\bibitem[15]{Cao}Y. Cao, I. D. Parker, G. Yu, C. Zhang, A.
J. Heeger, {\it Nature} {\bf 397} 414 (1998).

\bibitem[16]{Costa}V. C. Costa, M. J. Lochhead, K. L. Bray, {\it Chem. Mater.}
{\bf 8}, 783 (1996).

\bibitem[17]{Ribeiro}S. J. L Ribeiro, K. Dahmouche, C. A.
Ribeiro, C. V. Santilli, S. H. J. Pulcinelli, {\it J. Sol-Gel Sci.
Tech.} {\bf 13}, 427 (1998).

\bibitem[18]{Lianos99}V. Bekiari, P. Lianos, P. Judeinstein {\it Chem. Phys.
Lett.} {\bf 307}, 310 (1999).

\bibitem[19]{Carlos99c}L. D. Carlos, R. A. Sá Ferreira, V. De Zea Bermudez,
Celso Molina, Luciano A. Bueno, Sidney J.L. Ribeiro, {\it Phys.
Rev. B} {\bf 60}, 10042 (1999)

\bibitem[20]{Carlos00}L. D. Carlos, Y. Messaddeq, H. F. Brito, R. A. Sá Ferreira,
V. de Zea Bermudez, S. J. L. Ribeiro, {\it Adv. Mater.} {\bf 12},
594 (2000).

\bibitem[21]{Canham}L. T. Canham, {\it Appl. Phys. Lett.} {\bf 57}, 1046 (1990).

\bibitem[22]{Wilson}W. L. Wilson, P. F. Szajowski, L. E. Brus {\it Science}
{\bf 262}, 1242 (1993).

\bibitem[23]{Hirschman}K. D. Hirschman, L. Tsybeskov, S. P. Duttagupta,
P. M. Fauchet, {\it Nature} {\bf 384}, 338 (1996).

\bibitem[24]{Armand}M. Armand, C. J. Poinsignon, J.-Y. Sanchez,
V. De Zea Bermudez, U.S. Pat., {\bf 5,283,310} (1994).

\bibitem[25]{Bermudez92}V. De Zea Bermudez, D. Baril, J.-Y. Sanchez, M. Armand, C. J.
Poinsignon, in {\it Optical Materials Technology for Energy
Efficiency and Solar Conversion XI: Chromogenics for Smart
Windows}, A. Hugot-Le Golf, C.-G. Granqvist, C.M. Lampert, Eds.
(Proceedings SPIE, 1992), Vol. 1728, p. 180.

\bibitem[26]{Karim}K. Dahmouche, C. V. Santilli, S. H. J. Pulcinelli,
A. F. Craievich, {\it J. Phys. Chem. B} {\bf 103}, 4937 (1999).

\end{references}
\end{document}